\documentclass[a4paper]{article}
\usepackage[pdftex]{graphicx}
\pdfoutput=1
\usepackage{geometry}
\newcommand{\be}{\begin{equation}}
\newcommand{\ee}{\end{equation}}
\newcommand{\bd}{\begin{displaymath}}
\newcommand{\ed}{\end{displaymath}}
\newcommand{\bea}{\begin{eqnarray}}
\newcommand{\eea}{\end{eqnarray}}
\newcommand{\bi}{\begin{description}}
\newcommand{\ei}{\end{description}}
\newcommand{\bq}{\begin{quote}}
\newcommand{\eq}{\end{quote}}

\begin{document}
\bibliographystyle{unsrt}
\author{
Alexander~Unzicker\\
         {\small Pestalozzi-Gymnasium M\"unchen, Germany}\\
           {\small aunzicker@web.de}\\
Karl~Fabian\\
      {\small  Geological Survey of Norway, Trondheim, Norway}\\
   {\small karl.fabian@ngu.no }}
\title{Do Galaxies Change in Size~? An Angular Size Test at Low Redshift with SDSS data}
\maketitle

\begin{abstract}
Based on magnitudes and Petrosian radii from the Sloan Digital Sky Survey (SDSS, data release 7)
at low redhift ($z <0.2$), we developed a test of galaxy-size evolution. 
For this first quantitative size analysis using SDSS data, 
several possible sources of systematic errors had to be considered. 
The Malmquist bias is excluded by volume-limited samples.
A correction for seeing has been developed and applied.
We compare different methods to perform the  $K$-correction, and
avoid selection effects due to different filters.
It is found that apparent average galaxy size slightly decreases 
with redshift $z$, corresponding to a growth in time.
The effect is smaller for a lower $H_0$,
and at the same time less pronounced at higher redshifts,
but persists in both cases.
Although there is no systematic variation with galaxy luminosity, we
took into account the recently discovered luminosity evolution with redshift.
Assuming this effect of unknown origin to be real, we find a slight
increase of galaxy size with $z$.
The relative change of average size with $z$ usually amounts to less than one half
of the respective increase of wavelengths due to the cosmological redshift, making
a cosmological interpretation difficult. While the effect shows clear statistical 
significance, unknown systematics cannot be excluded. In any case, 
the enigmatic observations of size and luminosity evolution
need to be analysed together. To facilitate further investigations, a complete Mathematica 
code and instructions for data download are provided.

\end{abstract}

\section{Introduction}
The {\em Sloan Digital Sky Survey\/} (SDSS, data release 7) provides
free access to a rich source of galaxy data for a wide scientific community.
The number of topics  for which SDSS data are relevant is huge,
and the increasingly precise database facilitates
more and more detailed investigations.
The present study focusses on the
question whether galaxies at larger distances have the same average size and size distribution
as those in our neighborhood.
Such a connection between redshift and size has been considered
merely in in terms of cosmological models and galaxy evolution.
It is therefore useful to separate different redshift regimes.

\paragraph{High redshift.}
Being an important constraint to distinguish cosmological models,
several analyses of galaxy size have been published \cite{Nip:09, Ban:99, Sha:10, Mlc:10}.
The last reference \cite{Mlc:10} provides an excellent review of the relevant literature.
 A primary goal is
 to test whether deviations from the Euclidean relation $\theta \propto \frac{1}{z}$  between
  visual angle $\theta$ and redshift $z$ occur.
Such deviations are expected for the standard model based on Friedmann-Lemaitre
cosmology, according to which the minimum angular
size occurs at $z_{\min} \approx 1.5$ \cite{Rai:02, Har:02}.
Therefore, the natural distance to
distinguish different cosmologies is at high redshift $z > z_{\min}$, and
almost all investigations concentrate on this region.
A recent review of such observations concludes that
$\theta \sim \frac{1}{z}$ still is  compatible even with the data at high
redshift \cite{Mlc:10}. In particular, galaxies at redshift
$z=3.2$ appeared to have a six times {\em smaller\/} radius 
than predicted by $\Lambda$CDM.
This apparent discrepancy between  observation and the current cosmological
model is commonly interpreted in terms of galaxy evolution, which is assumed to influence size and luminosity.
Whether the necessary {\em growth\/} of galaxies with time is possible, is intensely
debated \cite{Nip:09, Fri:09, Wuy:10, Mlc:10}.
Because the mechanisms of galaxy evolution are not sufficiently well quantified,
this debate is still ongoing.
In any case, a galaxy evolution which is in agreement with $\Lambda$CDM,
must be most pronounced at high redshift, i.e. in the early universe.

\paragraph{Low redshift.}
To avoid the difficulties related to vigorous galaxy evolution at high redshift,
we here focus on galaxy observations at low redshift,
where galaxy evolution plays a minor role.
In this region the relation
$\theta \propto \frac{1}{z}$ should be in very good agreement with  observations,
because late-type galaxies are considered to be completely
virialized systems, and therefore should not further change in size.
Previous work, which included studies of size at small $z$
did not find noticeable deviations \cite{Gur:99, Nai:10, She:03, Tak:99}.
On the other hand, SDSS is an ideal database to
test a possible size evolution at low redshift, because the
majority of the $\approx$ 700000 galaxies,
where spectra have been measured, have redshifts around $z=0.1$.
At this low redshift, not only
the minor role of galaxy evolution simplifies the analysis, but also other problems, like
magnitude and color corrections, are easier to handle.
The SDSS data set thus provides a huge, but clean sample with remarkably small
errors.

\paragraph{Overview of the article.}
Our statistical study of the SDSS data indicates
that average galaxy sizes change in time for small redshifts as well.
Therefore, a careful analysis of selection effects
is needed. In  section~\ref{sec:methods}, we try to enumerate  all possible  sources of bias to the data, and discuss
the techniques to account for them.
These include volume-limited sampling, a correction for seeing,
$K$-correction (for color shifts), and removing of the color-dependence of the Petrosian radius.
In the results section we report the size change with redshift and the conditions
under which it occurs. The discussion of the results and possible interpretations
are found in section~\ref{sec:discussion}.

\section{Methods}\label{sec:methods}
\subsection{Data selection and general approach}
To analyse a redshift-size-relation, one has to select a redshift range considering several conditions.
Large $z$-intervals would be helpful in order to identify a trend, but unfortunately
also the  effects of possible systematic errors increase.
Data reliability is best for small values of  $z$, but the largest number of SDSS  galaxies are located around $z=0.1$.
To cover all possiblities, we investigated $z$ ranges
from $\Delta z=0.02$ to $\Delta z=0.08$, centered around small redshifts
from $z=0.06$ to $z=0.14$. In the Mathematica code provided, all these values can be varied with little
effort.

The SDSS catalogue contains magnitudes in five color bands,
 centered at different wavelengths.
To determine magnitude and size we primarily used the central
$r$ filter around $\lambda \approx 623$~nm.
It typically has small errors,
and also the global selection criteria in SDSS  use
this $r$-band. In addition, the  $K$-correction, discussed below,
also requires to use one of the central filters.
However, to correct for systematic variations of size and magnitude with color,
we also took into account the neighboring $g$ band.
The  SDSS DR7 data set contains photometric information of more than $2.5$ million galaxies,
and for about 700000 objects the redshift $z$ was determined with spectra.
Because a precise $z$ determination is essential for our analysis,
we only use the latter set of galaxies.
Only a very small fraction of this sample is more distant than  $z=0.2$.
To obtain data with small errors, we followed the recommendations for a  `clean photometry for galaxies'
\cite{DR7photo}. Details on the flags we used are found in the
appendix.

The very first step in our analysis is to correct the redshift to
the rest frame of the cosmic microwave background \cite{Lin:96},
to account for the solar system motion $369$~~km/s 
towards the constellation Crater at
$\delta= -7.22^{\circ}, \alpha= 167.99^{\circ}$.
This results in adding a redshift of the order 0.001 to most of the galaxies.

\subsection{Determining distance and size of galaxies}

Redshift $z$ is proportional to distance only approximately for low
$z$. To determine the precise absolute distance which is needed to
calculate galaxy size we use the standard
expression for the angular diameter distance\footnote{It distinguishes
from the commoving distance by a factor $\frac{1}{1+z}$.} in the $\Lambda$CDM model
(see e.g. \cite{Mlc:10}) is given by
\begin{equation}
d_A(z) = \frac{c}{H_0} \frac{1}{1+z}
\int_{0}^{z} \frac{dx}{\sqrt{\Omega_M (1+x)^3+ \Omega_\Lambda}} \; , \label{dist}\end{equation}
where $H_0$ is the Hubble constant, $\Omega_M$ the density of matter
and  $\Omega_{\Lambda}$ the density of dark energy. The luminosity distance
instead, taking into account the correction for the Tolman surface brightness
relation (see \cite{San:01}), is given by
\be
d_L(z)= (1+z)^2 d_A(z).
\ee

\begin{figure}[hbt]
\includegraphics[width=9.0cm]{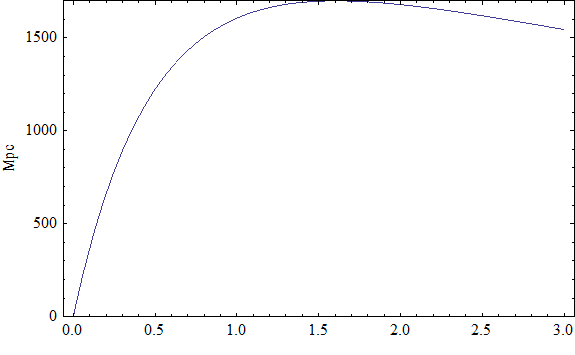}
\caption{Angular-diameter distance of astronomical objects as a function of $z$, according to
the concordance model with $\Omega_{\Lambda}=0.7$, $\Omega_{M}=0.3$, and $H_0= 72 \
km/s/Mpc$. Due to formally superluminal expansion velocities in the early
universe, the angular diameter distance peaks at $z \approx 1.5$ and becomes
smaller at higher $z$.}
\label{Edist}
\end{figure}

\begin{figure}[hbt]
\includegraphics[width=10.0cm]{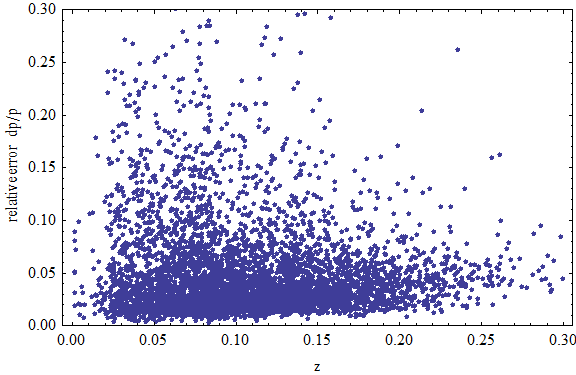}
\caption{Ratio of the error of the Petrosian radius and the
radius itself (r band), plotted for a random subset of the data. Only data points
below 0.2 were considered.}
\label{petroErr}
\end{figure}

\paragraph{The Petrosian radius.}
For the desired redshift-size relation, a measure of size is needed, but
unfortunately galaxies do not have sharp edges.
Therefore, sizes are commonly given in terms of the Petrosian radius
\cite{Pet:76}. It is defined as the radius
at which the surface brightness decreases to a given fraction
of the average surface brightness \cite{DR7, Bla:01}. By slightly modifying the
 original definition, SDSS uses a value of $20 \%$  \cite{DR7}.
Depending on galaxy models, the Petrosian radius contains a fixed
fraction of the total luminosity of the galaxy.
This is called the Petrosian magnitude
which is considered in the following.
To avoid a dependence on distance, which is model-dependent,
the Petrosian radius in SDSS is given in arcsec.
The automatic Petrosian-radius determination encounters various
difficulties, such as multiple radii or measurements
at faint surface brightness, which
are labeled by corresponding flags in the data\footnote{See the SQL
query in the appendix. Instead of using the `nopetro' flag which is sensitive
to all filters, we just took out the galaxies where a determination in
the $r$ and $g$ filters failed.}. To avoid any pathologic
behavior, we remove all those special cases from the analysis.
Additionally, we require the error in Petrosian radius not to exceed
$20 \%$ of its value.\footnote{For a large number of galaxies, the error
of the Petrosian radius is set to $50 \%$ of the radius, for the $u$ filter,
even to $100 \%$. }
As fig.~\ref{petroErr} shows, this seems to be a reasonable choice 
to exclude possible outliers, while keeping 
the bulk of the data available for the evaluation.
Altogether, we used redshift, extinction-corrected Petrosian magnitudes and Petrosian radii
in three filters and the corresponding errors so far. A detailed
description how to obtain the data is given in the appendix.

\subsection{The problem: Selection without selection effects}
\paragraph{Faint magnitude limit in SDSS and Malmquit bias.}
The Petrosian magnitude corrected for galactic extinction
is particularly important because it is used to
define the overall sensitivity of the database. Above (fainter as) the
limiting value $m_r=17.77$ mag in the $r$-band\footnote{$u,g,r,i,z$, are
centered at wavelengths $\lambda= 354 \ nm, 477 \ nm, 623 \ nm, 762 \ nm, 913 \ nm$.
The infrared filter $z$ has nothing to do with redshift.},
 only  few galaxies are found,
whereas the catalogue is considered to be complete below that limit.
For statistical studies, a tighter of $m_r=17.5$ 
is recommended; we followed that recommendation.

\begin{figure}[hbt]
\includegraphics[width=12.0cm]{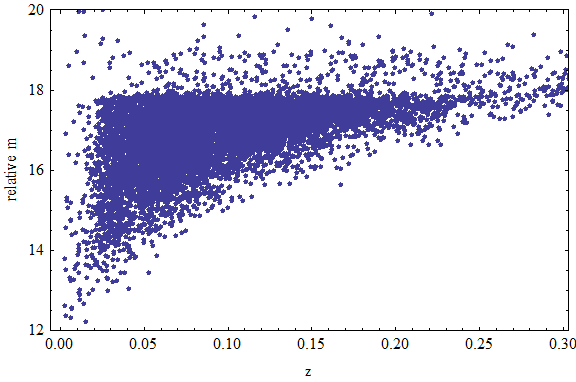}
\caption{Only a few galaxies with low
luminosity above the limit $m_r=17.77$ are visible. At large distances,
high luminosities are rare due to the distance. The plot shows a random subset
of the 600879 galaxies in consideration. }
\label{allG}
\end{figure}

The most prominent source of selection bias for astronomical
objects is the Malmquist bias. At larger distances, faint
galaxies go undetected, and since fainter usually means smaller,
they simply drop out of the sample.  
The situation is outlined in fig.~\ref{allG}.
Consequently, when  investigating a given redshift range, it makes no sense to
include at small $z$ a galaxy which would be invisible at larger
 $z$ due to the limit $m_r < 17.5$.

\begin{figure}[hbt]
\includegraphics[width=12.0cm]{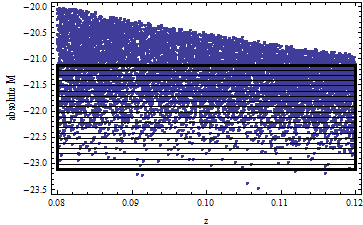}
\caption{Visualisation of the volume limited samples method. In each
of the horizontal stripes, only galaxies of the same absolute magnitude
are considered. Slice thickness could vary, here $dM=0.2$ mag. A random subset
of all data points is plotted.}
\label{vleplot}
\end{figure}

\paragraph{Volume limited samples.}
To avoid the Malmquist bias due to luminosity, we
implemented the following method: the galaxy sample
with different magnitudes at different redshifts
(see fig.~\ref{vleplot}) is subdivided into `stripes' containing
galaxies of the same {\em absolute\/} magnitude $M$. In each stripe,
galaxy size can be plotted as a function of $z$ (see later fig.~\ref{fitextwo}). 
Because there is no prior knowledge besides the equal $M$
for all those galaxies, the Malmquist bias is eliminated and
no size variation with $z$ should be expected so far. 

Although the faint stripes in the upper `triangle' of fig.~\ref{vleplot}
($M> -20.65 \ mag$) in principle could be used, a corresponding analysis would
contain more data in the low-$z$ parts of the given redshift range
(here 0.08-0.12). Thus we consider only galaxies with
an absolute magnitude which is within the faint-magnitude limit of
$m_r=17.5$~mag at the maximal redshift (here 0.12). On the other hand,
saturation effects make luminosities brighter than $m_r=14.5$~mag
unreliable. This corresponds to an absolute magnitude at the
minimal redshift, which should be excluded from the analysis for
analogous reasons. Therefore, in the
chosen $z$ range (volume), {\em all\/} galaxies in the corresponding
magnitude range (see rectangle in fig.~\ref{vleplot}) are visible,
thus this is called a volume limited sample. As a consequence, a larger
redshift range $\Delta z$ leads to a smaller range in absolute
magnitude $M$ and vice versa.

\begin{figure}[hbt]
\includegraphics[width=10.0cm]{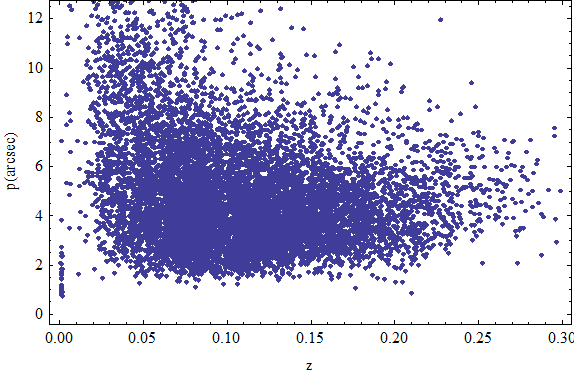}
\caption{Petrosian radii (r band) in arcsec plotted against redshift z,
for a random subset of the main galaxy sample. The star-galaxy-cut at $p \approx 2$ arcsec is
visible.}
\label{petrodis}
\end{figure}

\paragraph{Angular size selection effects.}
Whilst the volume-limited-sample method avoids the
unwanted brightness-selection effects, additional
caution has to be exercised when analyzing sizes, i.e.
Petrosian radii of galaxies. Due to the target selection algorithm
there is a necessary cut in angular size between stars and
galaxies, and very few Petrosian
radii lie below 2 arcsec (see fig.~\ref{petrodis}).
Without any precaution, this would  bias  
the results, because at large distance, 2 arcsec correspond to a 
larger galaxy size than at close distance. To be sure, we remove
therefore all  galaxies from the data set, which {\em would\/} appear at a
smaller angle than 2.2~arcsec {\em at the maximum redshift\/}.
A similar procedure is applied by \cite{She:03} to the
smaller quantity $petroR50$.\footnote{$R50$ however is still
more affected by seeing than the Petrosian radius \cite{Bla:03a}, fig.4.}
The numerical value of the cut
can be varied as a parameter in our code.
It corresponds to a cutoff below a certain absolute size in kpc
for the whole sample. Consequently, we also define
an overall {\em upper\/} limit for galaxy size (about 20 kpc)
to avoid data points with huge errors.
Thus, analogous to the rectangular form of volume limited
samples in a redshift-magnitude diagram, we additionally
applied a corresponding rectangle in a redshift-size diagram
for our analysis. Any pathology arising from improper
selection should be avoided by these methods.

\paragraph{Density and luminosity anomalies.}
It should be noted that, although all care has been  exercised
while selecting the data, the density of galaxies still does not correspond
precisely to the naive assumption of a mean constant density at large scales.
However, since it is generally established that the 
main galaxy sample is complete exceeding
$99 \%$ \cite{Str:02, Mal:09},
this cannot influence  the results observed here.\footnote{\cite{Bla:03}
describes the distribution of galaxies by a density parameter.}

Independent of the problem investigated here, it was recently
 found that galaxy luminosity clearly increases
with $z$ \cite{Bla:01,Bla:03, Lov:04}. This is in principle consistent
 with models of of stellar
evolution, although a quantitative understanding is still missing.
 Usually, the effect is described by an evolution parameter, 
 determined by \cite{Bla:03} to be $1.62 \ mag$ per unit redshift
for the $r$ band . Thus, in our code we
allow for this luminosity evolution, and include
1.62 as a variable parameter.

\subsection{K-correction}
The discussion so far is based on the assumption that
the magnitudes in the respective filters $u,g,r,i,z$ are
comparable at different redshifts. Unfortunately, this is
not true, since light of a distant galaxy which was originally
say in the $g$ or even $u$ filter, due to the Hubble redshift is  detected in
the $r$ filter.\footnote{E.g., a redshift $z=0.3$ would shift the
$g$ center $477 \ nm$ to the $r$ center $623 \ nm$.}
 This distance-dependent effect needs  careful treatment, called
$K$-correction, and various groups studied in detail
how the filter magnitudes transform into the rest
system $z=0$ \cite{Hog:02, Bla:03a}. We use the $kcorr_r$ values from the photo $z$ table in SDSS.
An approximation of similar
quality, but based on a simpler technique, is described in \cite{Chi:10},
where a fifth-order polynomial in  $z$ and $g-r$
(difference of magnitudes in the $g$ and $r$ filter)  nicely
reproduces the more detailed analysis. Since it is easily accessible,
the polynomial approximation is implemented in our code as well (see appendix).

\subsection{The impact of seeing on the Petrosian radius}

\begin{figure}[hbt]
\includegraphics[width=10.0cm]{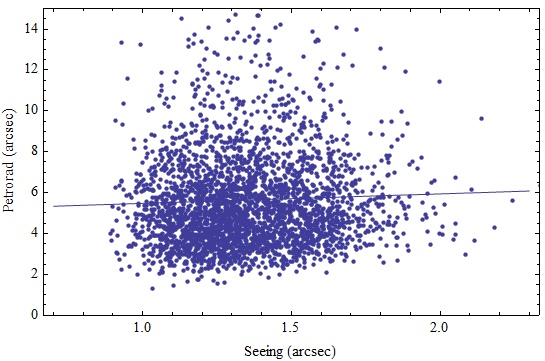}
\caption{The effect of seeing on the Petrosian angle, exemplarily for the
$r$ filter. The slope of the linear fit is $0.465$, while for the
$g$ filter $0.502$ is obtained. The undisturbed size of a galaxy can
be recovered by extrapolating to seeing 0. A random subset is plotted.}
\label{seeing}
\end{figure}

Another considerable problem for obtaining reliable Petrosian radii
is the effect of seeing.
It is obvious that bad atmospheric conditions tend
to smear out
galaxy profiles. This is dangerous in principle, because the
relative effect should be more pronounced at smaller angles and larger
distances. Limited seeing could therefore mimic a size increase with redshift,
as already noted in \cite{Str:02} (fig.~4). Unfortunately, seeing
affects all angular-size measures.
This even has been shown  for the
galaxy-light-concentration factor $c=\frac{p_{90}}{p_{50}}$ \cite{Bla:02}
\footnote{$p_{90}$ and $p_{50}$ denote the radii (in arcsec)
where the respective percentage of the Petrosian magnitude is
found (the Petrosian magnitude is defined by the light within {\em two} Petrosian
radii).}. Whilst $p_{90}$ is less dependent,  every
angular-distance measure tends to increase with seeing $s$.
To determine the relation between galaxy radius $p$ and seeing $s$, we fit all pairs $(s,p)$
(see fig.~\ref{seeing}) by a linear function, which yields a best-fit
 slope of about $0.5$ for the $r$ and $g$ filters.
Thus, the `true' Petrosian angle can be estimated
individually by extrapolating to perfect seeing $s=0$.
To test the dependence of the above seeing correction on outliers,  
we bin the data into intervals of width
0.1 in seeing $s$. The medians of these bins yield  17 data points for $0.7 < s < 2.4$.
Those were weighted by the number of galaxies and again fitted by a linear
function. We obtained 0.462 for the slope in $r$ and 0.483 for the slope in $r$,
a marginal difference to the above values (see fig.\ref{seeing}).

\subsection{The impact of color on the Petrosian radius}

\begin{figure}[hbt]
\includegraphics[width=10.0cm]{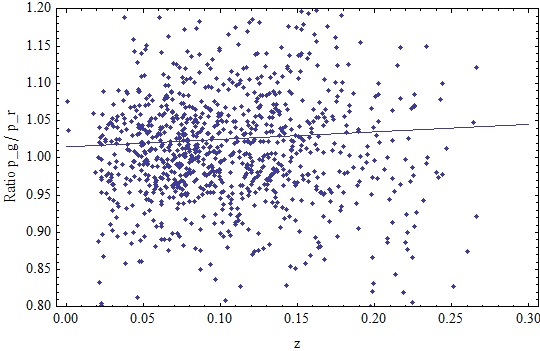}
\caption{The ratio of the Petrosian radii in the $g$ and $r$ band is a function
of redshift. The effect is corrected by the linear fit.}
\label{grcomp}
\end{figure}

Besides the  dependence on seeing, one also must
consider the influence of redshift. The more prominent effect
on magnitude is accounted for by the K-correction, but there
could be an impact of the redshift on the Petrosian radius
if the galaxy has different radii in different color bands. 
The idea to correct for this effect is to use $z$ to determine a linear interpolation of the
 $g$ and the $r$ radius which is independent of $z$.
First however, one has to assure that $p_g$ and $p_r$ are on average of
the same size. After having corrected for seeing,
we analyzed the ratio $p_g/p_r$ over redshift (fig.~\ref{grcomp})
and found a slight dependence:
\be
\frac{p_g}{p_r}= 1.0145+0.10 z
\ee
Taking this into account, the interpolation could be calculated as follows:
Since redshift $z=0.306$ would transform the center of the
$g$ filter ($477 nm$) to the center of the $r$ filter ($623 nm$),
our redshift dependent radius was computed as
\be
p(z)=p_g + \frac{z}{0.306} (p_r - p_g).
\ee
To  ensure using only absolutely reliable data,
 we remove all galaxies from the data set where $p_g$ and $p_r$  differ by more than
$20 \% $.

\paragraph{Cosmological parameters.}
Though we expressed our results in terms of redshift, angular diameter distances
for the galaxies were necessary to correct both for the absolute magnitude
and for the computation of the real size from the angular (Petrosian) radius.
The  distances (\ref{dist})
depend on the Hubble constant $H_0$ and on the densities of matter $\Omega_M$
and dark energy $\Omega_{\Lambda}$. All quantities can be varied in our
program as parameters. Thus the galaxy size $R$ in kpc is simply
\be
R=\frac{2 \pi d_A(z) p_r}{3600*360},
\ee
where $p_r$ is the Petrosian angle in arcsec (r-band) and the absolute magnitude
is
\be
M= m_r + 5 \lg (d_L(z))- K(z, m_g-m_r) - 25,
\ee
where $m_r$, $m_g$ are the extinction-corrected magnitudes in the u,g,r,i,z filters
and $K(z, m_g-m_r)$ is the K-correction of \cite{Chi:10}, while `individual' K-corrections from
SDSS are considered as well.

\subsection{Further selection criteria and data reduction}
Several  conflicting effects
have to be balanced when selecting appropriate parameters
for our investigation: 
1) a larger range $\Delta z$ makes it
easier to detect possible trends. 2)  the number
of systematic effects and their possible errors increases
for large $\Delta z$, e.g. the K-correction.
3) as  evident from fig.~\ref{vleplot}, a large range $\Delta z$
leads to a small range in magnitudes and many galaxies are cut off
by the volume limited sample method.

$\Delta z$ can be used as a variable parameter and  
the largest numbers of galaxies in the data set occur for $0.02 < \Delta z < 0.08$.
The number of galaxies is also used as a guiding principle for
choosing the $z$ location of the galaxy sample. The peak of
our original magnitude-limited ($m_r=17.5$) population is located at
$z \approx 0.09$. Thus we concentrate our analysis on the interval
$z=0.06$ to $z=0.14$ where most of the data points lie.
Another parameter to choose is the thickness in magnitude
of the `stripes' $dM$
used to divide the volume limited samples as in fig.~\ref{vleplot}.
Given that peculiar velocities are in the range of $1000 \ km/s$ 
which corresponds to a uncertainty $\delta z \approx 0.0033$,
this leads to an error of almost 0.1 mag at $z=0.10$.
Therefore, we choose  $dM =0.02$ as default value.

\begin{figure}[hbt]
\includegraphics[width=15.0cm]{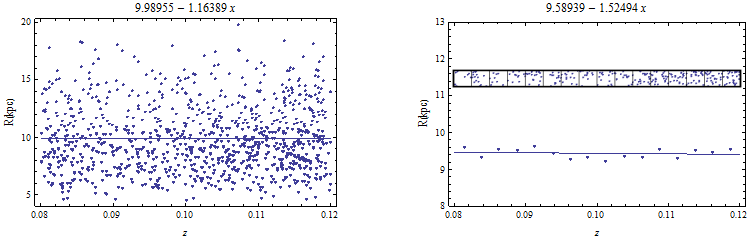}
\caption{Left: one stripe from fig.~\ref{vleplot}
with absolute magnitude $-21.2 < M < -21.0$ is considered and the size is plotted.
Though of the same magnitude, galaxies show a large scatter in size. Nevertheless, a trend of
size change can be described by a linear function. The slope $kpc/dz$ and the
extrapolation to $z=0$ is shown. Right:
The stripes at a fixed magnitude in fig.~\ref{vleplot} are now divided into
small boxes in $z$. Among all galaxies in one box, the median of the size is considered.
Again, the trend can be fitted with a linear function.}
\label{fitextwo}
\end{figure}

\paragraph{A First approach: Linear fit of size trends.}
We  exemplarily look at
{\em one\/} stripe $0.08 < z < 0.12$ and $ -20.7 > M > -20.9$
selected by the volume-limited-sample method described above
(fig.~\ref{vleplot}).
Though having the same absolute magnitude, the sizes of
the galaxies differ considerably.  The large scatter
is illustrated in fig.\ref{fitextwo} (left).
A linear least-square fit
to the data points\footnote{We required
a minimum number of 300 galaxies in each stripe.}
yields a slope (in kpc/redshift), and an $R$-axis intercept,
which can be interpreted as the average radius of a galaxy
with the given luminosity at $z=0$.

\paragraph{Improved fit of bin medians.}
As visible in fig.~\ref{fitextwo} (left), the scatter in size
for galaxies of the same magnitude is considerable and may give rise to line-fit
errors. Since the median is not sensitive to
possible outliers, instead of fitting the data directly, as in
fig.~\ref{fitextwo} (left), we first calculate the median of the Petrosian radii within small intervals
$dz =0.0025$, as shown in fig.\ref{fitextwo} (right), and then fit a line to these medians.
This procedure reduces a data set like fig.\ref{fitextwo} (left) to
$\frac{\Delta z}{d z} = 8 $ points and avoids the otherwise implicit
weighting by the number of galaxies that varies with redshift. There is however 
another subtlety that justifies this procedure. Due to the overall size cut to avoid
outliers, e.g. only galaxies with $4.5 \ kpc < R < 20 \ kpc$ are considered for $0.08 < z < 0.12$.
While this is appropriate for an average luminosity, a significant number of bright galaxies  
are cut by $R < 20 \ kpc$ while the distribution of faint galaxies is affected by the cut $R > 4.5 \ kpc$.
In this case, the mean radius overestimates the real value for faint galaxies, while the
mean radius of the bright galaxies is distorted towards smaller values. For both effects,
taking the median instead of the mean is the appropriate remedy. It is further clear
the direct fit without median would underestimate any size trend  (see Table~1). 
Thus we consider the median fit to be the cleaner procedure than fitting directly all
data points in fig.~\ref{fitextwo} (left). 
Such a linear median fit is computed for every magnitude `stripe'.
The slope of these functions
is then a measure of size increase with $z$.

\begin{figure}[hbt]
\includegraphics[width=15.0cm]{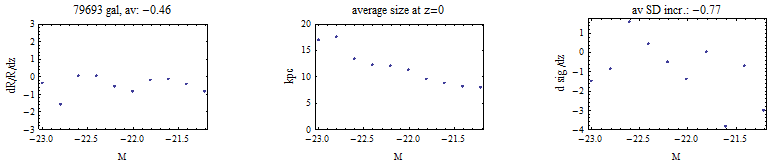}
\caption{Typical result of our analysis: relative increase $\frac{dR}{R_0 dz}$
of galaxy size with redshift for different magnitudes $M$ (left), characteristical
average size for different magnitudes (middle) and increase of the standard
deviation of the size distribution (right). 79693 galaxies considered in the
from redshift $z=0.08$ to $z=0.12$, K-correction from SDSS, 2.2 arcsec and
$20 \ kpc$ cut, fit of medians, no evolution, reference size $R_0$ from
individual fit extrapolated to $z=0$.
$R_0$ and $\frac{dR}{R_0 dz}$ are anti-correlated for single data points.}
\label{exres}
\end{figure}

But how to compare a size increase $dR/dz$ for galaxies with
different sizes? Therefore, we choose as meaningful
quantity the {\em relative\/} increase per unit redshift,
$\frac{dR}{R_0 dz}$. It will be of central importance in our results.
However, a meaningful value for the reference radius $R_0$ at
 redshift $z=0$ has yet to be found.
Though an individual linear fit yields a slope and
an  intersection estimating  $R_0$, the latter value, being an extrapolation, can have a large error.
As it can be observed in a typical result like fig.~\ref{exres}, a smaller 
intersection $R_0$ leads to a higher slope and vice versa; we seeked an $R_0$ that
avoided such an anticorrelation.

\paragraph{Determining a characteristic size-magnitude relation $R_0(M)$ for galaxies.}

To determine such a characteristic relation
between magnitude and average size, we ran the above algorithm 
for various redshift
intervals, resulting in a ensemble of $R_0(M)$ estimates, as shown in fig.~\ref{exres}
(middle). Then we fit all these estimates by the 3-parameter (a,b,c) {\em nonlinear\/} 
function(see later fig.~\ref{magSize}):
\be
R(M)=a \exp(b M + c M^2). \label{para}
\ee
The characteristic size-magnitude relation $R(M)$ obtained in this way provides a 
reasonable yardstick for subsequent
runs of the algorithm, where  the central
quantity $\frac{dR}{R dz}$ now refers to this $R(M)$ 
(see fig.~\ref{magSize} below). Technically, it is of advantage to use
a more precise characteristic size-magnitude relation determined at $z=0.1$, 
instead to the extrapolated radii at $z=0$. In this case, we however had to apply 
a correction for the on average larger radii at $z=0.1$.

\paragraph{Properties of the size distribution.}

As an additional test, we were interested if the size distributions
of galaxies at different redshift showed a suspicious behaviour. E.g.,
a narrower distribution with increasing median would indicate an
artificial cut of a population of galaxies. A typical result is shown
in fig.~\ref{exres} (right).

\section{Results} \label{results}

\begin{figure}[hbt]
\includegraphics[width=15.0cm]{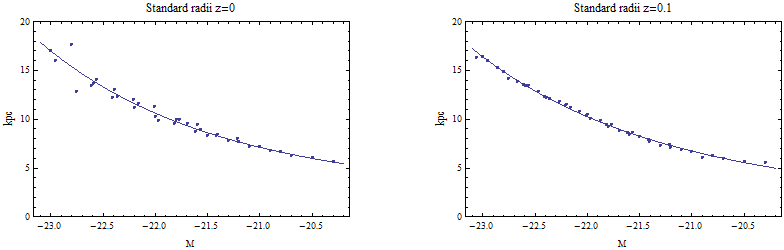}
\caption{Size-magnitude relation for galaxies at the rest system $z=0$, estimated from the
fit of linear function. size-magnitude relation at $z=0$ and $z=0.1$ derived from
five fits between $z=0.04$ and $z=0.16$ with a $z$ range of $\Delta z=0.04$. The lower
$z$ values were weighted for extrapolating to $z=0$, while at $z=0.1$ a much larger
number of galaxies leads to  increased precision.}
\label{magSize}
\end{figure}

\paragraph{Characteristic galaxy sizes $R_0$.}
The relation between average size and magnitude is obtained by
the 3-parametric fit described
in the last section. For the parameters in
(\ref{para}) we find for $z=0$ $\ln a = 17.67$, $b=1.808$, $c= 0.05055$
and $\ln a = 12.48$, $b=1.335$, $c= 0.03973$ for $z=0.1$.
Altogether, we found the following dependence
fig.~\ref{magSize}. Given the uncertainties, a quite consistent function
may be drived from $-20.2 > M > -23.1$ (depending on $H_0$, here $72 \ km/s/Mpc$).

\begin{figure}[hbt]
\includegraphics[width=15.0cm]{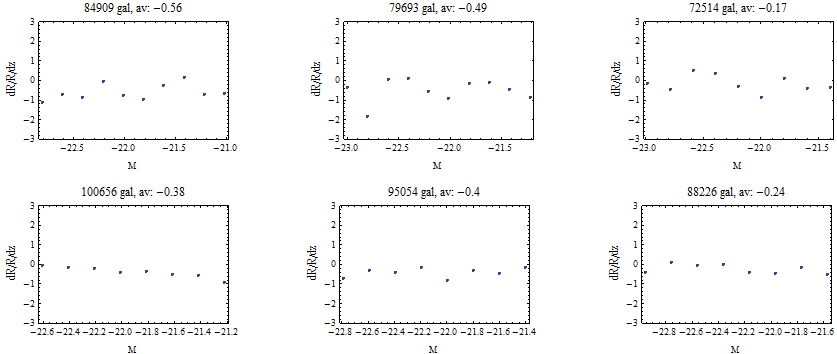}
\caption{Relative increase of galaxy size $dR/R$ per redshift $dz$ as a function of absolute magnitude $M$.
Redshift center from left to right: $0.09, 0.10, 0.11$. Redshift range from top to bottom:
$0.04; 0.06$. Slices $0.2$ mag, K-correction from SDSS Photo z table, $p> 2.2$ arcsec, size limit $20 \ kpc$.}
\label{mainresK}
\end{figure}

\begin{figure}[hbt]
\includegraphics[width=15.0cm]{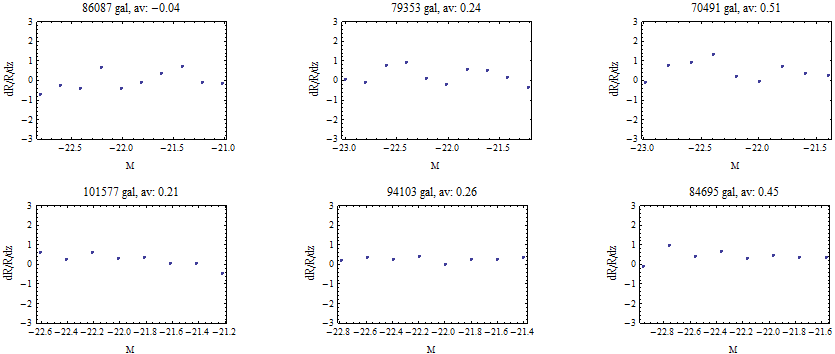}
\caption{As fig.~\ref{mainresK}, but now considering luminosity evolution:
relative increase of galaxy size $dR/R$ per redshift $dz$ as a function of absolute magnitude $M$.
Redshift center from left to right: $0.09, 0.10, 0.11$. Redshift range from top to bottom:
$0.04, 0.06$. Slices $0.2$ mag, K-correction from SDSS Photo z table.}
\label{mainresKE}
\end{figure}

\paragraph{Main result: galaxy size change with $z$.}
The results of our redshift-size analysis are shown in figs.~\ref{mainresK}-\ref{mainresKE},
each small picture for a different redshift regime.
The negative values of $\frac{dR}{R dz}$ indicate an average size
decrease with redshift $z$, equivalent to a growth in time.
The trend is less pronounced at higher redshift but occurs at all magnitudes.
While in fig.~\ref{mainresK}
the average of $\frac{dR}{R dz}$ over all magnitudes is given, one could think
about weighting. Since the number of galaxies decreases dramatically with magnitude,
weighting by the number would lead to faint galaxies  dominating the result.
As a compromise, often used in statistics, a square-root-weighted average is also
considered, all these quantities are displayed in the summary Table~1. It is
quite natural that the few bright galaxies show a relatively larger
scatter (see fig.~\ref{exres} right). 
While fig.~\ref{mainresK} refers to the K-correction provided by SDSS, 
we repeated our
analysis with a simple polynomial expression for the K-correction given by  \cite{Chi:10} depending
on $z$ and filter magnitudes only. Thereby, the average of 
$\frac{dR}{R dz}$ is slightly increased (see summary of results). 
The remarkable difference is that SDSS provides
positive values for the correction without exception, while the polynomial expression by \cite{Chi:10}
yields a considerable number of negative values at small redshifts.
Another type of K-correction depending on the $r$ and $u$ filter (instead of $r$ and $g$)
results only in insignificant changes (not shown here), while without K-correction
(a physically unmotivated case) the effect is masked. 
The effect of color due to different redshifts on the value of the Petrosian
radius turned out to be negligible.

\paragraph{Taking into account luminosity evolution.}

Given the findings of \cite{Bla:03} on the redshift dependence of the
luminosity function\footnote{The distribution of galaxies over the range of luminosities
is usually fitted with a Schechter function.}, we were also interested whether
our effect could be understood as a consequence of it. Instead of taking stripes
of equal luminosity in fig.~\ref{vleplot}, we were considering a sample of galaxies with
increasing luminosity in z (-1.62 mag per unit z in the r-band \cite{Bla:03}).
However, since magnitude and size are correlated, this led to a selection of brighter
and therefore bigger galaxies at higher redshift. Thus the average size increases now with $z$,
correspinding to a shrinking in time, as shown in fig.~\ref{mainresKE}. 
It seems to be difficult to explain within the common picture of galaxy evolution.

\begin{figure}[hbt]
\includegraphics[width=15.0cm]{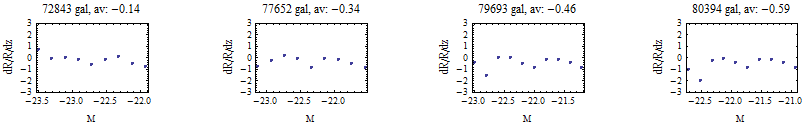}
\caption{As fig.~\ref{mainresK}, but with different Hubble
constants $H_0= 52, 62, 72, 82 \ km/s/Mpc$ (from left to right). The number
 of galaxies and the median of $\frac{dR}{R dz}$ is shown (fit without median).}
\label{hubble}
\end{figure}

\paragraph{Cosmological parameters.}
As expected, the results depend on the Hubble constant, though quite moderately
see Table~1. and fig.~\ref{hubble}. Additionally, we varied $\Omega_M$ from $0.22$
to $0.32$ while keeping $\Omega_M+\Omega_{\Lambda}=1$ fixed, with a still
smaller effect than for varying $H_0$.
Plotting all possible parameter variations and their combinations would require excessive
space. 
Most of the applied correction methods had an impact on the results.
 Cutting out the rectangular  volume limited samples
 from fig.~\ref{vleplot} led to smaller values of $\frac{dR}{R dz}$,
 and so did the consideration of the angular size limits.

\clearpage

\paragraph{Seeing.}
The cutoff value in the Petrosian angle due to the star-glaxy separation turned out
to be significant in principle, but the value of 2.2 arcsec carefully chosen.
Changing it to 2.7 arcsec decreased significantly the number of galaxies
in the analysis, but not did not influence the results very much.

\begin{figure}[hbt]
\includegraphics[width=15.0cm]{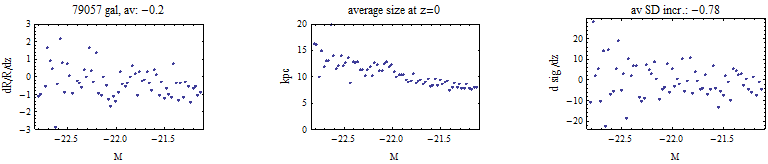}
\caption{As fig.~\ref{exres}, but with artificially
small magnitude stripes of $dM=0.025$ mag. Since the error in absolute magnitude
is at least twice as much due to peculiar velocities only, single data points are
subject to considerable statistical scatter. Remarkably,
the median of $\frac{dR}{R dz}$ is still in the same range.}
\label{ErrDemoM}
\end{figure}

\paragraph{Statistical errors.}
In view of the clear significance of the effect we did not perform a detailed
statistical analysis. Rather it is illustrative to
demonstrate the impact of a large statistical scatter on
our results. To introduce noise, it suffices to choose parameters obviously
outside a reasonable ranage. E.g., the thickness of the magnitude stripes could
be chosen much inferior to the typical error in magnitude due to peculiar velocities
(see fig.~\ref{ErrDemoM})
%

\newpage
\paragraph{Summary of results.}
Based on the detailed results displayed in fig~.\ref{mainresK}-\ref{mainresKE},
we give summary in table~1. The influence of absolute magnitude M is now
included in different ways of averaging $\frac{d R}{R dz}$.

\vspace{1.0cm}
\begin{tabular}{|lccc|ccc|}
\hline
{\bf z range} & 0.07-0.11 & 0.08-0.12 & 0.09-0.13 & 0.06-0.12 & 0.07-0.13 & 0.08-0.14\\
\hline
{\bf Default (fig.~\ref{mainresK})} &&&&&&\\
\hline
average & -0.56 & -0.49 & -0.17 & -0.38 & -0.40 & -0.24\\
sqrt-weighted av. & -0.50 & -0.46 & -0.23 & -0.46 & -0.38 & -0.29\\
weighted av. & -0.48 & -0.46 & -0.28 & -0.52 & -0.36 & -0.33\\
\hline
{\bf K-corr. polynomial:} &&&&&&\\
\hline
average & -0.61 & -0.58 & -0.24 & -0.45 & -0.5 & -0.38\\
sqrt-weighted av. & -0.54 & -0.53 & -0.27 & -0.52 & -0.47 & -0.38\\
weighted av. & -0.50 & -0.53 & -0.31 & -0.57 & -0.45 & -0.40\\
\hline
{\bf No K-correction:} &&&&&&\\
\hline
average & -0.07 & 0.26 & 0.46 & 0.20 & 0.14 & 0.2\\
sqrt-weighted av. & -0.04 & 0.21 & 0.34 & 0.07 & 0.09 & 0.28\\
weighted av. & -0.03 & 0.16 & 0.26 & -0.03 & 0.05 & 0.28\\
\hline
{\bf $H_0 = 52 $:} &&&&&&\\
\hline
average & -0.30 & -0.16 & -0.02 & -0.34 & -0.16 & -0.06\\
sqrt-weighted av. & -0.36 & -0.3 & -0.08 & -0.40 & -0.22 & -0.09\\
weighted av. & -0.39 & -0.38 & -0.12 & -0.47 & -0.22 & -0.12\\
\hline
{\bf Fit without median:} &&&&&&\\
\hline
average & -0.30 & -0.18 & -0.02 & -0.24 & -0.13 & -0.05\\
sqrt-weighted av. & -0.36 & -0.23 & -0.02 & -0.27 & -0.18 & -0.06\\
weighted av. & -0.38 & -0.26 & -0.02 & -0.31 & -0.2 & -0.07\\
\hline
{\bf Cut at 2.7 arcsec:} &&&&&&\\
\hline
average & -0.54 & -0.48 & -0.22 & -0.38 & -0.44 & -0.22\\
sqrt-weighted av. & -0.47 & -0.44 & -0.29 & -0.45 & -0.43 & -0.27\\
weighted av. & -0.42 & -0.43 & -0.34 & -0.49 & -0.43 & -0.31\\
{\bf Evolution (fig.~\ref{mainresKE})} &&&&&&\\
\hline
average & -0.04 & 0.24 & 0.51 & 0.21 & 0.26 & 0.45\\
sqrt-weighted av. & 0.05 & 0.20 & 0.46 & 0.12 & 0.26 & 0.45\\
weighted av. & 0.09 & 0.15 & 0.41 & 0.04 & 0.26 & 0.43\\
\hline
z range & 0.07-0.11 & 0.08-0.12 & 0.09-0.13 & 0.06-0.12 & 0.07-0.13 & 0.08-0.14\\
\hline 
\end{tabular}
\vspace{1.0cm}

Table~1.\\
Relative increase of galaxy size per unit redshift, $\frac{d R}{R dz}$. Average taken over
different magnitudes, weigthed and sqrt-weighted with the number of galaxies, corresponding
to fig.~\ref{mainresK}. Default refers to: K-correction from SDSS photo z table, fit of medians,
no luminosity evolution, angular cut at $p < 2.2$ arcsec, size limit $20 \ kpc$.

\vspace{0.5cm}

\section{Discussion}
\label{sec:discussion}
We developed a method to analyze for galaxy-size evolution at low redshifts. 
We find a slight decrease
of average galaxy size with redshift, corresponding to a growth in time. This result does not
depend on galaxy luminosity, indicating that the various corrections applied 
were reasonable. The fact that this decrease is less pronounced at higher redshifts
is more difficult to interpret and may be due bias from the K-correction. 
However, none of the different K-correction methods tested made the anomaly disappear. 
The same holds for different values of $H_0$. While a smaller $H_0$,
corresponding to an older universe, yields a smaller change in size, the
effect does not disappear. The fit without taking the median and the cut at $2.7 \ arcsec$
slightly masks a change for the reasons given above. It is however very interesting
to see that the trend in size change is reversed when taking into account the 
luminosity evolution \cite{Bla:03}. Because there is obvious physical explaining
the luminosity evolution, we cannot decide which of the two puzzling effects, 
size or luminosity change, is real. While luminosity increase 
could originate from stellar processes, a change in size is more difficult to
understand. In any case, future analyses should consider both effects.

With respect to other results regarding size evolution,
our finding of a slight increase in time would correspond 
to the observation of  too small galaxies at very high redshift (e.g. \cite{Mlc:10}), though
a quantitative agreement cannot be deduced yet. Looking
at fig.~\ref{Edist}, it is clear that those results challenge the
angular-size-redshift-relation of the $\Lambda CDM$ model in 
particular at high redshift. It is also clear that such
an effect is less pronounced  at low redshifts where our analysis took place.

A pragmatic approach would be to introduce an independent parameter
describing physical processes leading to the observed growth.
Methodologically, this is dangerous because for galaxies we only have a limited number
of observable quantities: redshift, number density,
luminosity and size. On the other hand, we observe an anomalous density,
a luminosity evolution and unexpected changes in size. It is not evident
how a comprehensive understanding of these effects can be obtained within standard cosmology.

\section{Outlook}
We have here developed a quantitative method to identify
galaxy size evolution at small redshift. Our results present 
yet another riddle for the study of galaxies.
 We hope that our published code will facilitate further
 investigations of this effect.

\paragraph{Acknowledgement.}
We are grateful to Francesco Sylos Labini and Martin Lopez-Corredoira for helpful hints
and thank Simon Staude for assistance.
A.U. thanks for the comments of David Hogg, Tom Shanks, Rudi Schild, Vladimir Sokolov
and Rick Watkins during the conference `New directions in modern cosmology' and
and the Lorentz Center in Leiden for the hospitality.

\newpage

\appendix{{\Large{Source codes}}

\paragraph{SQL query.}
With the commands given below, all the data used in our analysis can be downloaded
from the SDSS site $http://cas.sdss.org/astro/en/tools/search/sql.asp$.
By taking away the top 20 constraint in the first row, the search will however
produce a timeout due to the exeeding of the SDSS row limit of 100000 lines.
Therefore, the z range has to be split up in different queries. A good idea
is to choose small $\Delta z$ ranges, typically 0.01 or even smaller. Check if there
is no timeout error, and save all the files in one directory without renaming them.
A Mathematica routine how to join the files again is given below.

\begin{verbatim}
-- this indicates a comment.
-- top 20 is just for a check. It has to be taken out later
select top 20 s.ra, s.dec, s.z as redshift, s.zconf,
(p.petroMag_u - p.extinction_u) as mag_u,
(p.petroMag_g - p.extinction_g) as mag_g,
(p.petroMag_r - p.extinction_r) as mag_r,
p.petroRad_g, p.petroRad_r,
p.petroRadErr_g, p.petroRadErr_r,
p.petroR50_g, p.petroR50_r,
p.petroR90_g, p.petroR90_r,
r.seeing_g, r.seeing_r,
h.kcorr_g, h.kcorr_r,
h.absMag_g, h.absMag_r
from galaxy p, specObj s, RunQA r, Photoz h
where p.objID = s.bestObjID and
p.fieldID = r.fieldID and
p.objID = h.objID and
-- s.specClass=2 and
s.z BETWEEN 0.0001 AND 0.02   --to be adjusted in steps:  0.03, 0.035, 0.04, 0.045...0.06,0.064,0.068,
-- 0.072, 0.076, .....0.10,0.105, ...0.
AND p.objID <> 0
AND (p.petroMag_r - p.extinction_r) < 17.77 -- faint magnitude limit for MGS
AND ((flags_r & 0x10000000) != 0)
-- detected in BINNED1
AND ((flags_r & 0x8100000c00a0) = 0)
-- not NOPROFILE, PEAKCENTER, NOTCHECKED, PSF_FLUX_INTERP, SATURATED,
-- or BAD_COUNTS_ERROR.
-- if you want to accept objects with interpolation problems for PSF mags,
-- change this to: AND ((flags_r & 0x800a0) = 0)
AND (((flags_r & 0x400000000000) = 0) or (psfmagerr_r <= 0.2))
-- not DEBLEND_NOPEAK or small PSF error
-- (substitute psfmagerr in other band as appropriate)
AND (((flags_r & 0x100000000000) = 0) or (flags_r & 0x1000) = 0)
-- not INTERP_CENTER or not COSMIC_RAY - omit this AND clause if you want to
-- accept objects with interpolation problems for PSF mags.
-- AND ((flags_r & 0x0000000000800000) = 0) -- petrofaint
-- AND ((flags_r & 0x0000000000000100) = 0) -- nopetro
AND ((flags_r & 0x0000000000000400) = 0) -- nopetro_big
-- AND ((flags_r & 0x0000000000000200) = 0) -- manypetro
AND ((flags_r & 0x0000000000002000) = 0) -- manyr50
AND ((flags_r & 0x0000000000004000) = 0) -- manyR90
AND ((flags_r & 0x0000000100000000) = 0) -- DEBLENDED_AS_MOVING
AND ((flags_r & 0x0000000000400000) = 0) -- badsky
order by s.z

\end{verbatim}

\paragraph{Mathematica code - preliminaries.}

The following commands which work irrespective of the names given
to the downloaded files produce a datafile of the type we used. The
CMB correction is also calculated here. You need to name your working
directory accordingly. Writing several files of about 100 MB and the
CMB calculation may need considerable time up to 30 mins. This has to be
done only once, however.

\begin{verbatim}
Needs["VectorAnalysis`"];(* glueing files with different ranges to one file:
store your SDSS datafiles like result (13).csv in a separate subdircetory named sdssgals*)
mydir ="c:\\Users\\sascha\\Desktop\\sdss\\";(* replace this with your math dir*)
SetDirectory[mydir <> "sdssgals"];
li = FileNames[];
compl = {{"ra", "dec", "redshift", "zconf", "mag_u", "mag_g", "mag_r",
    "petroRad_g", "petroRad_r", "petroRadErr_g", "petroRadErr_r",
   "petroR50_g",  "petroR50_r", "petroR90_g",  "petroR90_r",
   "seeing_g", "seeing_r", "kcorr_g", "kcorr_r", "absMag_g",
   "absMag_r"}}; For[kk = 1, kk <= Length[li], kk++,
 wer = Drop[Import[li[[kk]], "CSV"], 1];
 AppendTo[compl, wer]]; out = Flatten[compl, 1];
SetDirectory["c:\\Users\\sascha\\Desktop\\sdss"];
(*Export["allgal.csv",out,"CSV"];*)out >> "allgal.txt";
allGalaxies = Drop[<< "allgal.txt", {1, 21}];
(* CMB correction: 10 min, for that stored in separate file*)
CMBShift[x_] :=Block[{dis, halb},
dis = CoordinatesToCartesian[{1, Pi/2 - Pi (x[[2]])/360, Pi (x[[1]])/360}, Spherical] -
CoordinatesToCartesian[{1, Pi/2 - Pi 7.22/360, Pi 167.99/360}, Spherical];
halb = ArcTan[Sqrt[Plus @@ (dis^2)]/2];
zadd = Round[0.00123*Cos[2 halb], 0.000001]];
xx = OpenWrite["allGalCMB2.txt"];
For[ii = 1, ii <= Length[allGalaxies], ii++, linie = allGalaxies[[ii]];
 add = CMBShift[linie];
 linie3 = ReplacePart[linie, {3 -> linie[[3]] + add}];
 WriteString[xx, linie3[[3]], " ", linie3[[4]], " ", linie3[[5]], " ",
   linie3[[6]], " ", linie3[[7]], " ", linie3[[8]], " ", linie3[[9]],
  " ", linie3[[10]], " ", linie3[[11]], " ", linie3[[16]], " ",
  linie3[[17]], " ", linie3[[18]], " ", linie3[[19]], " ",
  linie3[[20]], " ", linie3[[21]]];
 Write[xx]]; Close[xx];
\end{verbatim}

\paragraph{Mathematica code - main analysis.}

The first paragraph still contains preliminaries that need not to be run
every time. At the very first run, the comment (*.. *) has to be dropped in
line 12-14 in order to produce the file galBuff.txt, which is smaller
and can be used in the following.

\begin{verbatim}
mydir = "c:\\Users\\sascha\\Desktop\\sdss";(* put your working directory here*)
SetDirectory[mydir]; Needs["Combinatorica`"]; Needs["ANOVA`"];
Needs["StatisticalPlots`"];
cc = 299792.458; minmag = 17.5; maxmag = 14.5;(* speed of light and mag range*)
xq = Table[{}, {20}];(*contains graphics*)
LimitedSample[lst_, lim_] :=
 Select[lst, (#[[lim[[1]]]] >= lim[[2]] && #[[lim[[1]]]] <=lim[[3]]) &];
LimitedSample2p[lst_, lim1_, lim2_] :=
Select[lst, (#[[lim1[[1]]]] >= lim1[[2]] && #[[lim1[[1]]]] <=
      lim1[[3]] && #[[lim2[[1]]]] > lim2[[2]] && #[[lim2[[1]]]] <= lim2[[3]]) &];
(*allGalaxies=Import["allgalCMB2.txt","Table"];tu=TimeUsed[];*)
(*vorgal=LimitedSample2p[allGalaxies,{5,maxmag, minmag+0.27},{2,0.9, 1.0}];
gal=LimitedSample2p[vorgal,{8,0,5},{9,0,5}]; gal>>"galBuff,txt";*)
(** starting with fainter than 17.5, otherwise kcorr diluites distribution*)
gal = << "galBuff.txt";
tgoR = Transpose[gal];
SeeAndPetg = Transpose[{tgoR[[10]], tgoR[[6]]}];
seeFit =Fit[SeeAndPetg, {1, x}, x]; psightg = seeFit[[2, 1]];
SeeAndPetr = Transpose[{tgoR[[11]], tgoR[[7]]}];
seeFit = Fit[SeeAndPetr, {1, x}, x]; psightr = seeFit[[2, 1]];
tgoR = Drop[Insert[tgoR, tgoR[[6]] - tgoR[[10]] psightg, 6], {7}];
tgoR = Drop[Drop[Drop[Insert[tgoR, tgoR[[7]] - tgoR[[11]] psightr, 7], {8}], -2], {10, 11}];
gal2 = Transpose[tgoR]; (* not everything is needed*)
grRatio = 1.0149; grSlope = 0.10095;
(*grpetrotest=Transpose[{tgoR[[1]],tgoR[[6]]/tgoR[[7]]}];
Fit[grpetrotest,{1,x},x]*)
(*** K-correct Polynomials Chilingarian et al. 2010*)
rWithgr = {{0, 0, 0, 0}, {-1.61166, 3.87173, -3.87312,
    2.66605}, {8.48781,
    13.2126, -6.4946, -7.31552}, {-87.2971, -35.0474, 41.5335,
    0}, {271.64, -26.9081, 0, 0}, {-232.289, 0, 0, 0}};
rWithur = {{0, 0, 0, 0}, {-1.98173, 1.04346,
    0.0221613, -0.0391318}, {9.34198, 1.639, -0.392805,
    0.192349}, {-39.8237, -10.3007, -1.9142, 0}, {123.94, 25.7117, 0,
    0}, {-150.964, 0, 0, 0}};
koeff = Table[c^i z^j, {j, 0, 5}, {i, 0, 3}];
KcorrRgr[c_, z_] = Plus @@ Flatten[rWithgr koeff];
KcorrRur[c_, z_] = Plus @@ Flatten[rWithur koeff];
\end{verbatim}

The following input defines the main routine Petroplot. All parameters can be varied here.

\begin{verbatim}
(*cosmological parameters, mag range, absolute mags considered, z range, minimum number of galaxies*)
PetroPlot[{H0_, Om_, OL_}, magstep_, {minz_, maxz_, dz_}, {minpetro_, maxsize_},
   minnumber_, {petroErr_, petroRatio_}, kflag_, distflag_, Rflag_, fitflag_, Epar_] :=
  Block[{zselect, zselectK, pselect, goodRad, seeingcorr},
   tu1 = TimeUsed[]; (* v1 corrected*)
   EmmissionDistInt2[z_] :=1/(1 + z) cc/H0 NIntegrate[1/(OL+(1+x)^3*Om),{x, 0, z}];
   If[distflag == 1,EmmissionDist =Interpolation[Table[{z, EmmissionDistInt2[z]}, {z, 0, 5, .02}]],
    Clear[EmmissionDist]; EmmissionDist[z_] = z*cc/2/H0 ];
   DistCorrect[z_] := -5 Log[10, (1+z)^2 EmmissionDist[z]] - 25; (* v1 corrected*)
   AbsPetR[tg_] :=EmmissionDist[tg[[1]]]*1000 *((tg[[6]] + (tg[[7]]*grRatio - tg[[6]])*
          tg[[1]] (grSlope + 1/0.30608)) /3600) Pi/180 ;
(* Galaxy sizes in kpc: now considering the shift from the g-band to the r-band*)
   (* redshift .30608 would shift the center of g to the center of r
   grratio is accounts for the ration of average g/r radii*)
   zselect = LimitedSample[gal2, {1, minz, maxz}];
   (* selecting z range and sufficient seeing conditions *)
(*now substituting with reduced pretorad due to seeing *)
   Print["correcting for seeing with coefficients g,r: ", {psightg, psightr}];
   goodRad = Select[zselect, (1/petroRatio < #[[7]]/#[[6]] < petroRatio) &];
   pselect = Select[goodRad, ((#[[8]]/(#[[6]]) <
          petroErr) && (#[[9]]/(#[[7]]) <
          petroErr) && ((#[[6]] + (#[[7]]*grRatio - #[[6]])*#[[1]] (grSlope + 1/0.30608)) >
          minpetro*EmmissionDist[maxz]/
       EmmissionDist[#[[1]]]) && ((#[[6]] + (#[[7]]*grRatio - #[[6]])*#[[1]] (grSlope + 1/0.30608))*
       1000/3600*Pi/180*EmmissionDist[#[[1]]] < maxsize)) &];
(*dropping huge errors in petrorad*)
   (* taking out all galaxies that would appear at < minpetro at
   the maximum redshift, thus avoiding a size bias *)
   (*taking a linear combination of the radii in the r and g band*)
   Print[
    "Total sample/z+faint mag/ petro constraints: ", {Length[gal],
     Length[zselect], Length[pselect]}];
   Kcorr[c_, z_] :=
    Switch[kflag, 0, 0, 1, KcorrRgr[c, z], 2, KcorrRur[c, z]];
   usedData = {#[[1]],
       EmmissionDist[#[[1]]], #[[5]] + DistCorrect[#[[1]]] -
        If[kflag == -1, #[[11]],
         Kcorr[#[[5 - kflag]] - #[[5]], #[[1]]]] + (#[[1]] - 0.1)*
         Epar, AbsPetR[#]} & /@ pselect;
   (* only redshift, distance,
   luminosity and size in the following *)
   (* now accounting for evolution , Blanton et. al.2003:*)
   (* determination of reasonable magnitudes in the given z range *)
   minabs = minmag + DistCorrect[maxz];
   maxabs = maxmag + DistCorrect[minz];
   Print["Original Range: ", {minabs, maxabs}];
   slices =Table[Select[
      usedData, ((ii >= #[[3]]) &&  #[[3]] > ii - magstep) &], {ii,
      minabs, maxabs, -magstep}];
   lastslice = Mod[minabs - maxabs, magstep];
   (* take out the sets with a small galaxy number*)
   While[Length[slices[[1]]] < minnumber, slices = Delete[slices, 1];
    minabs -= magstep];
   count = 0;(*
   taking into account that the last slice could be smaller than magstep*)
   While[Length[slices[[-1]]] < minnumber,
    slices = Delete[slices, -1];
    maxabs += If[count == 0, lastslice, magstep]; count += 1;];
   mla = Map[Length, slices];
   mags = Take[Table[j, {j, minabs, maxabs, -magstep}] - magstep/2, {1,
      Length[mla]}];
   pairstab = Table[{Mean[#[[3]] & /@ slices[[i]]],
      Map[{#[[1]], #[[4]]} &, slices[[i]]]}, {i, 1, Length[slices]}];
   chest = Table[{pairstab[[i, 1]],
      Select[pairstab[[i,
        2]], (minz + k*dz < #[[1]] < minz + (k + 1)*dz) &]}, {k,
      0, (maxz - minz)/dz - 1}, {i, 1, Length[pairstab]}];
   mags = Table[chest[[i, k, 1]], {k, 1, Length[chest[[1]]]}, {i, 1,
      Length[chest]}];
   zMedi =
    Table[Median[Transpose[chest[[i, k, 2]]][[1]]], {k, 1,
      Length[chest[[1]]]}, {i, 1, Length[chest]}];
   newVari =Table[Sqrt[Variance[Transpose[chest[[i, k, 2]]][[2]]]], {k, 1,
      Length[chest[[1]]]}, {i, 1,
      Length[chest]}];
   newMedi =
    Table[Median[Transpose[chest[[i, k, 2]]][[2]]], {k, 1,
      Length[chest[[1]]]}, {i, 1, Length[chest]}];
   Medians =
    Table[{mags[[k, i]], {zMedi[[k, i]], newMedi[[k, i]]}}, {k, 1,
      Length[chest[[1]]]}, {i, 1, Length[chest]}];
   Variances =
    Table[{mags[[k, i]], {zMedi[[k, i]], newVari[[k, i]]}}, {k, 1,
      Length[chest[[1]]]}, {i, 1, Length[chest]}];
   newTabOfFits =
    Select[Table[{Medians[[j, 1, 1]],
       Fit[If[fitflag == 0, pairstab[[j, 2]],
         Transpose[Medians[[j]]][[2]]], {1, x}, x]}, {j,
       Length[Medians]}], NumberQ[#[[2, 1]]] == True &];
   newTabOfFitsV =
    Select[Table[{Variances[[j, 1, 1]],
       Fit[Transpose[Variances[[j]]][[2]], {1, x}, x]}, {j,
       Length[Variances]}], NumberQ[#[[2, 1]]] == True &];
   RelincR =
    If[Rflag == 0,
     Map[{#[[1]], #[[2, 2, 1]]/#[[2, 1]] } &, newTabOfFits],
     Map[{#[[1]], #[[2, 2, 1]]/SizeMag10[#[[1]]] } &, newTabOfFits]];
   RelincV =
    If[Rflag == 0, Map[{#[[1]], #[[2, 2, 1]] } &, newTabOfFitsV],
     Map[{#[[1]], #[[2, 2, 1]]} &, newTabOfFitsV]];
   R0 = Map[{#[[1]], #[[2, 1]] } &, newTabOfFits];
   R10 = Map[{#[[1]], #[[2]] /. x -> 0.1 } &, newTabOfFits];
   rp = ListPlot[R0, Frame -> True, Axes -> False,
     FrameLabel -> {"M", "kpc"}, PlotRange -> {0, 20}, Frame -> True,
     PlotLabel -> "average size at z=0"];
   weiAv = Round[Plus @@ ((#[[2]] & /@ RelincR)*mla)/Plus @@ mla, 0.01];
   sqrAv = Round[Plus @@ ((#[[2]] & /@ RelincR)*Sqrt[mla])/Plus @@ Sqrt[mla],0.01];
   avraw = Mean[Transpose[RelincR][[2]]];
   avV = Median[Transpose[RelincV][[2]]];
   av = Round[If[Rflag == 0, avraw, avraw/(1 - 0.1 avraw)], 0.01];
   sqav = Round[If[Rflag == 0, sqrAv, sqrAv/(1 - 0.1 sqrAv)], 0.01];
   weav = Round[If[Rflag == 0, weiAv, weiAv/(1 - 0.1 weiAv)], 0.01];
   Print[Plus @@ mla, " Galaxies of ", Length[pselect], " considered"];
   Print["in the absM range: ", {minabs, maxabs}];
   Print["Distribution: ", mla];
   Print["Weighted Average dR/R/dz: " , weav];
   Print["Sqrt-Average dR/R/dz: " , sqav];
   Print["average dR/R/dz: " , av]; tu2 = TimeUsed[];
   Print["time used: " , tu2 - tu1];
   pp = ListPlot[RelincR, Frame -> True, Axes -> False,
     FrameLabel -> {"M", "dR/R/dz"}, PlotRange -> {-3, 3},
     PlotLabel ->
      ToString[Plus @@ mla] <> " gal, av: " <> ToString[av]];
   pV = ListPlot[RelincV, Frame -> True, Axes -> False,
     FrameLabel -> {"M", "d sig /dz"}, PlotRange -> All,
     PlotLabel -> "av SD incr.: " <> ToString[Round[avV, 0.01]]]];
\end{verbatim}
Now, the code can be run and visualized with
\begin{verbatim}
PetroPlot[{72, 0.3, 0.7}, 0.2, {0.08, 0.12, 0.0025}, {2.2, 20}, 300, {0.2, 1.2}, -1, 1, 0, 1, 0];
Show[GraphicsArray[{pp, rp, pV}]]
\end{verbatim}
However, for our final resultes we used $Rflag=1$, which needs a function to be calculated
by the following procedure which stores the characteristic radii in a file. Afterwards,
Pertrorad can be repeated with Rflag=1 (third parameter from behind)
\begin{verbatim}
(** first step of determination of standardradii in the rest system Rflag=0*)
StandardRadii = StandardRadii10 = {}; For[i = 0, i <= 4, i++,
 PetroPlot[{72, 0.3, 0.7},
  0.2, {0.04 + 0.02 i, 0.08 + 0.02 i, 0.0025}, {2.2, 20},
  300, {0.2, 1.2}, -1, 1, 0, 1, 0]; Print[i];
 (*weighting where more galaxies are *)
 For[kk = 4, kk > i, kk--, AppendTo[StandardRadii, {R0, mla}]];
 For[kk = 4, kk >= (i - 2)^2, kk--,
  AppendTo[StandardRadii10, {R10, mla}]]];
{StandardRadii, StandardRadii10} >>"SRadiiK.txt";
(*or get it from data*)
{StandardRadii, StandardRadii10} = <<"SRadiiK.txt";
R0List = Flatten[#[[1]] & /@ StandardRadii, 1];
R10List = Flatten[#[[1]] & /@ StandardRadii10, 1];
(* function necessary to run with Rflag=1*)
SizeMag[m_] =
 Exp[Fit[{#[[1]], Log[#[[2]]]} & /@ R0List, {1, m, m^2}, m]];
SizeMag10[m_] =
 Exp[Fit[{#[[1]], Log[#[[2]]]} & /@ R10List, {1, m, m^2}, m]];
rlp = ListPlot[R0List, PlotRange -> {0, 20}, Frame -> True,
  PlotLabel -> "Standard radii z=0"]; smlp =
 Plot[SizeMag[m], {m, -22.7, -20.0}, PlotRange -> {0, 20},
  Frame -> True];
rlp10 = ListPlot[R10List, PlotRange -> {0, 20}, Frame -> True,
  PlotLabel -> "Standard radii z=0.1"]; smlp10 =
 Plot[SizeMag10[m], {m, -22.7, -20.0}, PlotRange -> {0, 20},
  Frame -> True];
g0 = Show[rlp, smlp]; g10 = Show[rlp10, smlp10];
xq[[5]] = Show[GraphicsArray[{g0, g10}], FrameLabel -> {"M", "kpc"}]
\end{verbatim}
Now, run again
\begin{verbatim}
PetroPlot[{72, 0.3, 0.7}, 0.2, {0.08, 0.12, 0.0025}, {2.2, 20}, 300, {0.2, 1.2}, -1, 1, 1, 1, 0];
\end{verbatim}
\end{document}